\documentclass[%
reprint,
superscriptaddress,
aps,
pra,
floatfix,
]{revtex4-2}

\usepackage{graphicx}
\usepackage{dcolumn}
\usepackage{bm}
\usepackage{systeme}
\usepackage{ulem}
\usepackage{xcolor}
\usepackage{amsmath} 
\usepackage{hyperref}
\usepackage{multirow}
\usepackage[
scale=0.9, a4paper, marginratio={1:1, 1:1}, ignoreall,
]{geometry}
\definecolor{newtext}{RGB}{0, 0, 0}

\begin{document}


\title{
Effect of magnetic anisotropy relaxation on laser-induced magnetization precession \\ in thin galfenol films
} 

\author{P. I. Gerevenkov}
\email{petr.gerevenkov@mail.ioffe.ru}
\homepage{http://www.ioffe.ru/ferrolab/}
 \affiliation{Ioffe Institute, 194021 St. Petersburg, Russia}
\author{D. V. Kuntu}
 \affiliation{Ioffe Institute, 194021 St. Petersburg, Russia}
 \author{Ia. A. Filatov}
 \affiliation{Ioffe Institute, 194021 St. Petersburg, Russia}
 \author{L. A. Shelukhin}
 \affiliation{Ioffe Institute, 194021 St. Petersburg, Russia}
  \author{M.~Wang}
 \affiliation{School of Physics and Astronomy, The University of Nottingham, Nottingham NG7 2RD, UK}
   \author{D.~P.~Pattnaik}
 \affiliation{School of Physics and Astronomy, The University of Nottingham, Nottingham NG7 2RD, UK}
 \author{A. W. Rushforth}
 \affiliation{School of Physics and Astronomy, The University of Nottingham, Nottingham NG7 2RD, UK}
\author{A. M. Kalashnikova}
 \affiliation{Ioffe Institute, 194021 St. Petersburg, Russia}
\author{N. E. Khokhlov}
 \affiliation{Ioffe Institute, 194021 St. Petersburg, Russia}

\date{\today}

\begin{abstract}

The rate and pathways of relaxation of a magnetic medium to its equilibrium following excitation with intense and short laser pulses are the key ingredients of ultrafast optical control of spins. 
Here we study experimentally the evolution of the magnetization and magnetic anisotropy of thin films of a ferromagnetic metal galfenol (Fe$_{0.81}$Ga$_{0.19}$) resulting from excitation with a femtosecond laser pulse. 
From the temporal evolution of the hysteresis loops we deduce that the magnetization $M_S$ and magnetic anisotropy parameters $K$ recover within a nanosecond, and the ratio between $K$ and $M_S$ satisfies the thermal equilibrium's power law in the whole time range spanning from a few picoseconds to 3 nanoseconds. 
We further use the experimentally obtained relaxation times of $M_S$ and $K$ to analyze the laser-induced precession and demonstrate how they contribute to its frequency evolution at the nanosecond timescale.     

\end{abstract}

\maketitle


\section{\label{sec:level1}Introduction}

For the operation of spintronic and magnonic devices, various relaxation processes following perturbation of the magnetic state by an external stimulus play a role as important as the excitation processes themselves.
For instance, precessional switching of the magnetization in spin valves driven by magnetic field or electric current pulses requires substantial damping of the precession~\cite{slonczewski-JMMM1996}, while low damping is desirable for spin-torque nano-oscillators~\cite{kiselev-Nat2003, demidov-natmat2012} and spin-wave logic devices~\cite{ney-APL2005}.
On the other hand, heat dissipation processes create constraints for high operation rates of devices such as spin-transfer torque magnetic random access memory, STT-MRAM~\cite{bhatti2017spintronics, zhang2018addressing}, and heat-assisted magnetic recording, HAMR~\cite{xu2012thermal}. 

Recent progress in the ultrafast control of magnetic media with femto- and picosecond laser pulses has put extra emphasis on studying and manipulating relaxation processes following strong and fast optical excitation.
The rate of cooling of the electronic, phononic, and spin systems defines the laser fluence and duration suitable for all-optical single-~\cite{Vahaplar-PRB2012, davies2020pathways, gorchon2016role} and multiple-shot~\cite{kichin2019multiple, Medapalli2017Multiscale, john2017magnetisation} switching of magnetization in metals.
Laser-driven precessional switching of non-thermal~\cite{stupakiewicz-Nat2017} and thermal~\cite{Davies-PRL2019, shelukhin_prAppl_2020} origins appears to be enabled by the fine balance between the magnetic damping and the life-times of the altered magnetic anisotropy state.
The latter works have highlighted the importance of understanding not only how the magnetic anisotropy responds to laser-induced heating at the sub-picosecond timescale, but also how quickly it relaxes to the equilibrium value~\cite{KimelNature:2004, vanKampen:PRL2002, ShelukhinPRB:2018, carpene-PRB2008, Bigot-PRL2002}.

Magnetic anisotropy parameters at equilibrium can be obtained with high precision using ferromagnetic resonance~\cite{farle-1998} or torque measurements~\cite{miyajima1976simple}.
Since these techniques are incompatible with time-resolved measurements, the evolution of magnetic anisotropy following ultrafast laser excitation is usually evaluated indirectly by monitoring laser-driven precession.
This, however, allows reliable tracking of magnetic anisotropy changes of large magnitudes only~\cite{Atoneche-PRB2010, Hashimoto-PRL2008, stupakiewicz-Nat2017, afanasiev-NatMater2021}. 
In the case of subtle changes, assumptions regarding the temporal evolution of the magnetic anisotropy have often to be made. 

In this article we study experimentally the timescale of recovery of the magnetic anisotropy parameters of thin films of a ferromagnetic metallic alloy galfenol following ultrafast optical excitation and governed by laser-induced heating. 
We focus our study on the timescales beyond the first picosecond after the excitation, which are highly relevant to the processes of laser-driven precessional switching of magnetization, magnetostatic waves etc.
By studying laser-induced changes of magneto-optical hysteresis loops, we deduce the temporal evolution of the saturation magnetization and of the cubic and uniaxial magnetic anisotropy parameters independently.
This enables us to confirm that during relaxation at the nanosecond timescale the ratio between the magnetic anisotropy and saturation magnetization satisfies the established power-law relation for thermal equilibrium.
We further apply the experimentally obtained timescales of magnetization and anisotropy relaxation to the analysis of the laser-driven magnetization precession.
This allows a more precise evaluation of the precession frequency and tracking of its changes.

The article is organized as follows. 
In Sec.\,II we describe the galfenol samples and two experimental techniques used to track laser-induced magnetization and magnetic anisotropy evolution and precession. 
Section\,III presents experimental results and discussion. 
In Sec.\,III\,A we analyze the temporal evolution of the magnetization and magnetic anisotropy and discuss the corresponding timescales. 
Section\,III\,B is focused on a discussion of the magnetization precession parameters. 
This is followed by conclusions, where we also discuss relevance of the results to a number of prospective applications.

\section{Experimental details}

\subsection{Samples}

The experiments are performed using films of galfenol with thicknesses $d =$ 5 and 10\,nm.
The films are epitaxially grown on (001)-GaAs substrates by magnetron sputtering as described elsewhere~\cite{parkes-SciRep2013}.
All films are capped with protective chromium and silicon dioxide layers with thicknesses of 2.5\,nm and 120\,nm, respectively.
According to x-ray diffractometry on similar films, the films are polycrystalline with linear crystallite sizes of about 12\,nm and misorientation of the crystalline axes of order of 1~degree~\cite{bowe2017magnetisation}.
Since the sizes of the laser spots in all experiments are much larger than the sizes of individual crystallites, from here on the parameters of the sample are given in the approximation of a single-crystal film.
Galfenol films grown on GaAs substrates exhibit intrinsic cubic magnetocrystalline anisotropy with easy magnetization axes along $<100>$ crystallographic directions.
Additionally, there is a growth-induced uniaxial anisotropy with the easy axis along the $[110]$ direction~\cite{atulasimha2011review, parkes-SciRep2013, DanilovPRB:2018}.
The films are in-plane magnetized at equilibrium.
Such galfenol films were shown to support all-optical excitation of a homogeneous long-living precession and propagating spin waves via laser-induced ultrafast changes of the magnetic anisotropy~\cite{Scherbakov:PhysRevAppl2019, kats2016ultrafast, khokhlov-PRAppl2019}.

\subsection{Experimental setups}

\begin{figure}
\includegraphics[width=1\linewidth]{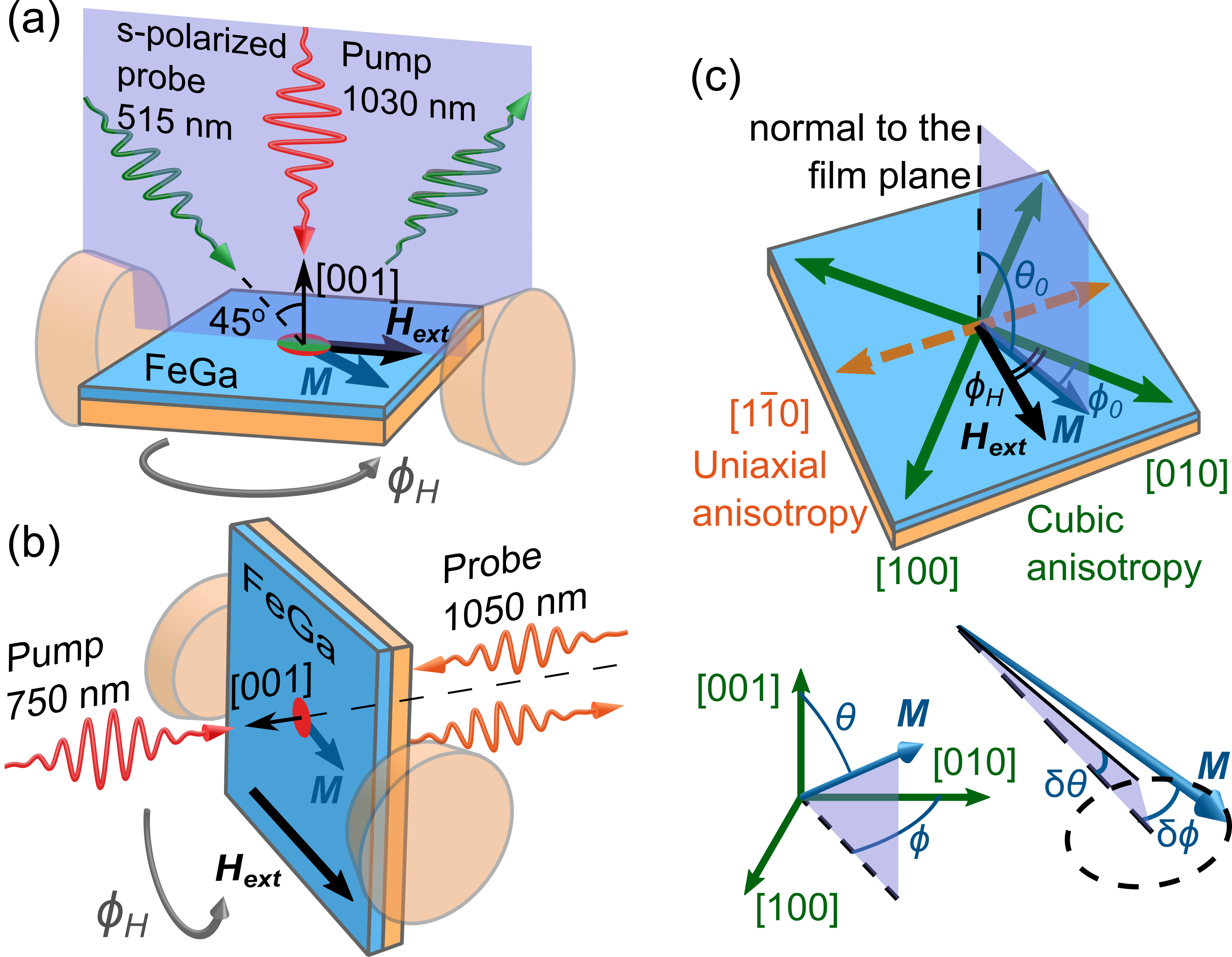}
\caption{\label{fig:geometry}The geometries of (a) TR-LMOKE, (b) TR-PMOKE experiments, and (c) definitions of angles used in the calculations.}
\end{figure}

We use two optical pump-probe setups with experimental geometries to measure the longitudinal and polar Kerr effects.
The setup of the time-resolved longitudinal magneto-optical Kerr effect (TR-LMOKE) is used to obtain the hysteresis loops at different time delays after the pump pulse excitation, and then to determine the temporal evolution of the saturation magnetization $M_S$ and the uniaxial $K_U$ and cubic $K_C$ anisotropy parameters, similar to works~\cite{roth2008dynamics, Li_coercivity_2013_JAP}.
The setup of the time-resolved polar magneto-optical Kerr effect \mbox{(TR-PMOKE)} is used to detect the magnetization precession after the pump pulse excitation.
The geometry of TR-LMOKE measurements is shown in Fig.~\ref{fig:geometry}\,(a).
The laser pulses with duration of 170\,fs, a central wavelength of 1030\,nm, and a repetition rate of 100\,kHz  generated by a Femtosecond regenerative amplifier are split into pump and probe.
The probe pulse is converted to the second harmonic with a $\beta$-BaB$_2$O$_4$ crystal.
Linearly polarized pump pulses are focused by lenses onto the film normally in a spot with diameter of 60\,$\mu$m.
S-polarized probe pulses are focused by lenses in a spot with diameter of 40\,$\mu$m at an angle of incidence of 45\,$^\circ$.
The measurements are performed at pump fluences $J =$ 7 and 14 mJ/cm$^2$.
The fluence of the probe is about 20 times lower.
Amplitudes of the pump pulses are periodically modulated at 997\,Hz with an optical chopper.
The polarization rotation angle $\beta_L$ of the reflected probe is detected using an optical bridge consisting of a Wollaston prism and a balanced detector connected to a lock-in amplifier synchronized with the chopper.
An external dc magnetic field is applied in the film plane, as shown in Fig.~\ref{fig:geometry}\,(a).
In the TR-LMOKE geometry, $\beta_L$ is proportional to the component of the magnetization ${\bf M}$, parallel to the external magnetic field $\mathrm{\bf H_{ext}}$.
The time delay between the pump and probe pulses is controlled with an optical delay line.
The hysteresis loops are measured as functions of $H_{ext}$ at various delay times $\Delta t$ between pump and probe pulses in the range of $-0.5..3$\,ns, where 0 corresponds to the simultaneous arrival of the pulses.

The geometry of \mbox{(TR-PMOKE)} measurements is shown in Fig.~\ref{fig:geometry}\,(b).
The pump and probe pulses with duration of 120\,fs, repetition rate of 70\,MHz, and central wavelengths of 750\,nm and 1050\,nm, respectively, are generated with an Femtosecond laser system with tunable wavelength.
Linearly polarized pulses are focused on the sample surface normally on the film side (pump) and through the substrate (probe) using micro-objective lenses into spots with diameters of 3\,$\mu$m.
The pump pulse fluence is 2\,mJ/cm$^2$.
The fluence of the probe pulses is about 20 times lower.
Amplitudes of the pump pulses are periodically modulated at 997\,kHz with an acousto-optical modulator.
The time evolution of the polarization angle $\beta_P$ is detected similarly to the method used in the TR-LMOKE setup.
In the TR-PMOKE geometry, $\beta_P$ is proportional to the out-of-plane component of magnetization $M_z$.
The measurements are performed at $\mu_0 H_{ext} = 100$\,mT applied in the film plane.
Here $\mu_0$ is the permeability of the vacuum.

The polarization of the pump pulses is not varied in the experiments.
Ultrafast changes of the magnetization $\bf M$ and the anisotropy parameters $K_U$ and $K_C$ are due to the laser-induced heating, which is polarization independent as discussed in detail elsewhere~\cite{kats2016ultrafast, ShelukhinPRB:2018, khokhlov-PRAppl2019}.
All measurements are performed at room temperature.

\section{Experimental results and discussion}

\subsection{\label{Sec:TemporalEvolution} Temporal evolution of magnetization and anisotropy parameters}

\begin{figure}
\includegraphics[width= \linewidth]{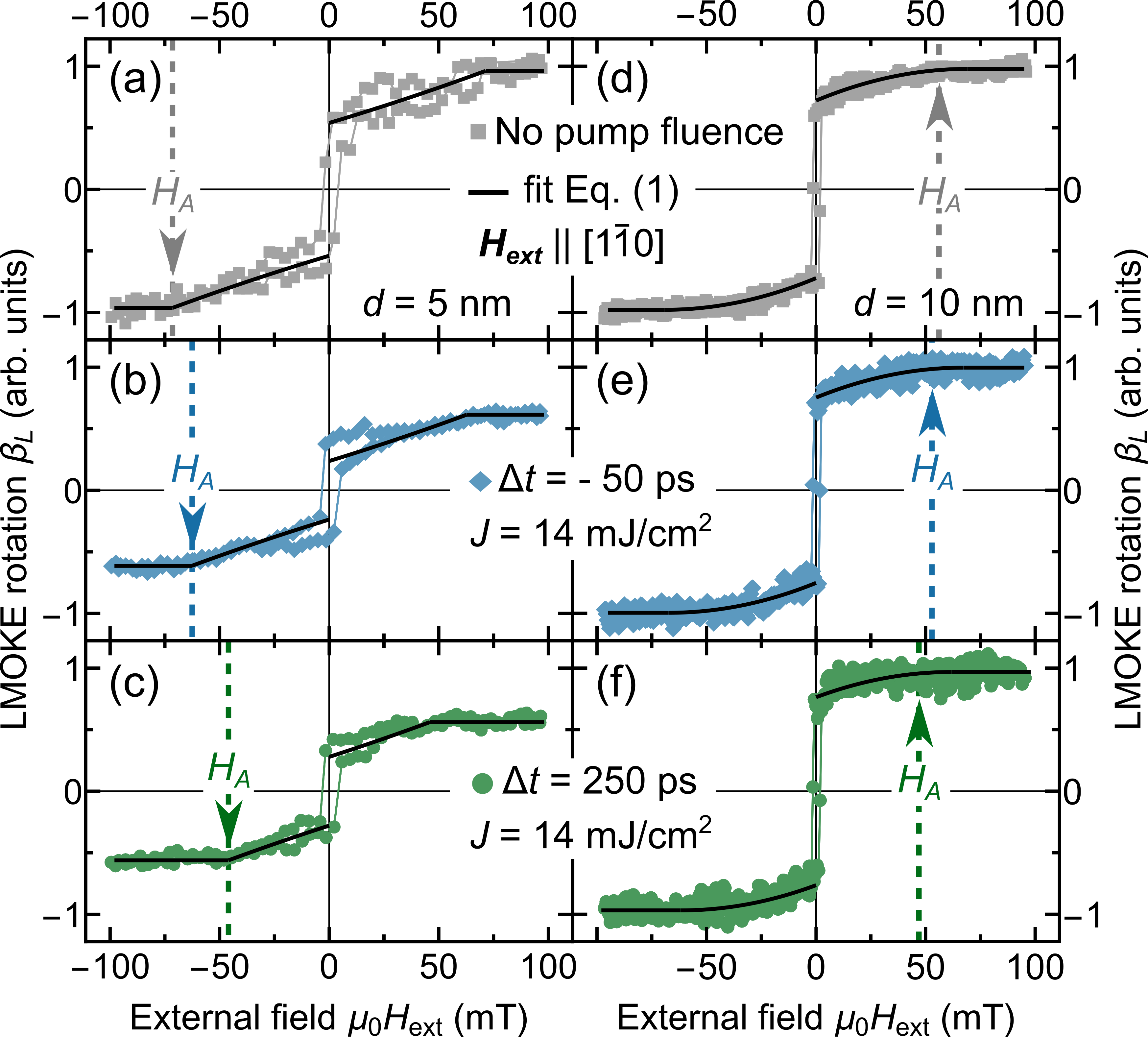}
\caption{\label{fig:loop} 
Field dependencies (symbols) of the LMOKE rotation $\beta_L$ as measured in the 5- and 10-nm-thick galfenol films without pump pulse excitation [(a), (d)] and under the excitation for two pump-probe delays $\Delta t$: (b), (e) before the excitation (-50\,ps) and (c), (f) at 250\,ps after the arrival of the pump pulse. 
The external field $\mathrm{H_{ext}}$ is parallel to $[1\overline10]$.
Solid lines show the fits with Eq.~(\ref{eq:Hist_fit}). 
The arrows show the corresponding values of the anisotropy field $H_A$ obtained from the fits. 
 }
\end{figure}

Figure~\ref{fig:loop} shows the typical field dependence of the LMOKE rotation obtained at ${\bf H_{ext}}$ along the hard magnetization axis ($HA$) $[1\overline10]$ for the 5 and 10-nm thick films.
One of the dependencies is obtained with no pump, while the others are obtained at $\Delta t = -50$ and 250\,ps.
All field dependencies possess the shape characteristic of the magnetization hysteresis loop of the material with cubic anisotropy when the field is applied along the HA.
In particular, the LMOKE rotation saturates at $\beta_{L}^{sat} \sim M_S$ when the field exceeds the value of the effective anisotropy field $H_A$ (marked with arrows in Fig.~\ref{fig:loop}).
Below this field LMOKE rotation exhibits monotonic growth with the field, and the abrupt switching in the vicinity of $H_\mathrm{{ext}}=0$. 
\textcolor{newtext}{The simplest expression which allows capturing the main features of the field dependencies in Fig.~\ref{fig:loop} for both films has the form}

\textcolor{newtext}{
\begin{equation}
\label{eq:Hist_fit}
\begin{split}
\beta_L(H_{ext}) = 
\begin{cases}
F(H_{ext}^{max} - H_{ext}),  & \text{if } H_{ext} > 0,\text{\;}\\
-F(H_{ext}^{max} + H_{ext}),  & \text{if } H_{ext} < 0,\\
\end{cases}\\
\text{where \qquad \qquad \qquad \qquad \qquad \qquad \qquad \qquad \quad}\\
F(x) = \beta_{L}^{sat} - \Theta\left( x - \xi \right) \left[ a_1 (x - \xi) + a_2 (x - \xi)^2 \right] 
\end{split}
\end{equation}
}

\textcolor{newtext}{\noindent where $H_{ext}^{max}$ is the maximum value of the external field in the loop, $\xi = H_{ext}^{max} - H_A$, $\Theta(x)$ is the Heaviside function, and $a_1$ and $a_2$ are the polynomial coefficients that provide good description of the experimental data for both 5 and 10-nm thick films. }
\textcolor{newtext}{The first term in $F(x)$ in Eq.~(\ref{eq:Hist_fit}) describes the saturation above $H_A$, while the second term accounts for the monotonic growth below $H_A$.}
\textcolor{newtext}{Details of the fit are given in Appendix~\ref{Sec:AppendixLoop}. }
As can be seen in Fig.~\ref{fig:loop}, impact of the laser pulse results in pronounced changes of the LMOKE field dependencies.
First, at both negative and positive time delays, there is a suppression of the signal amplitude $\beta_{L}^{sat}$, as well as a decrease of the saturation field $H_A$.
The laser-induced changes seen at the negative time delays can be readily ascribed to the accumulated heat occurring because of the laser excitation at high repetition rate.
Therefore, below we use this curve as a reference, representing the magnetic properties of the sample at an elevated, but stabilized, temperature. 

At positive delays there are two transient changes, i.e., those dependent on $\Delta t$.
The excitation by the pump pulse causes a transient decrease of $H_A$ and $M_S$.  
The temporal evolutions of $M_S$ and $H_A$ obtained from fitting the hysteresis loops with Eq.~(\ref{eq:Hist_fit}) at different $\Delta t$ and different orientation of $\mathrm{ \bf H_{ext}}$ allow one to determine the temporal evolution of the anisotropy parameters $K_U, K_C$.
Indeed, two different types of anisotropies (cubic and uniaxial) give the following form of the total anisotropy field $H_A$ along two different HA, $[1\overline{1}0]$ and $[110]$:

\begin{equation}
\label{eq:Ha}
\begin{split}
H_{A [1\overline{1}0]}(\Delta t) = H_{C}(\Delta t) + H_{U}(\Delta t),\\
H_{A [110]}(\Delta t) = H_{C}(\Delta t) - H_{U}(\Delta t),
\end{split}
\end{equation}
where $H_{C}$ and $H_{U}$ are the cubic and uniaxial effective anisotropy fields, respectively.

Equations~(\ref{eq:Ha}) allow us to separate the temporal dependencies of $K_C$ and $K_U$ by using the equations for anisotropy fields~\cite{chikazumi2009physics}:

\begin{equation}
\label{eq:K_vs_t}
\begin{split}
K_{U}(\Delta t) = \frac{1}{4} \mu_0 M_S(\Delta t) [H_{A [1\overline{1}0]}(\Delta t) - H_{A [110]}(\Delta t)],\\
K_{C}(\Delta t) = \frac{1}{2} \mu_0 M_S(\Delta t) [H_{A [1\overline{1}0]}(\Delta t) + H_{A [110]}(\Delta t)].
\end{split}
\end{equation}

\begin{figure}
\includegraphics[width=1 \linewidth]{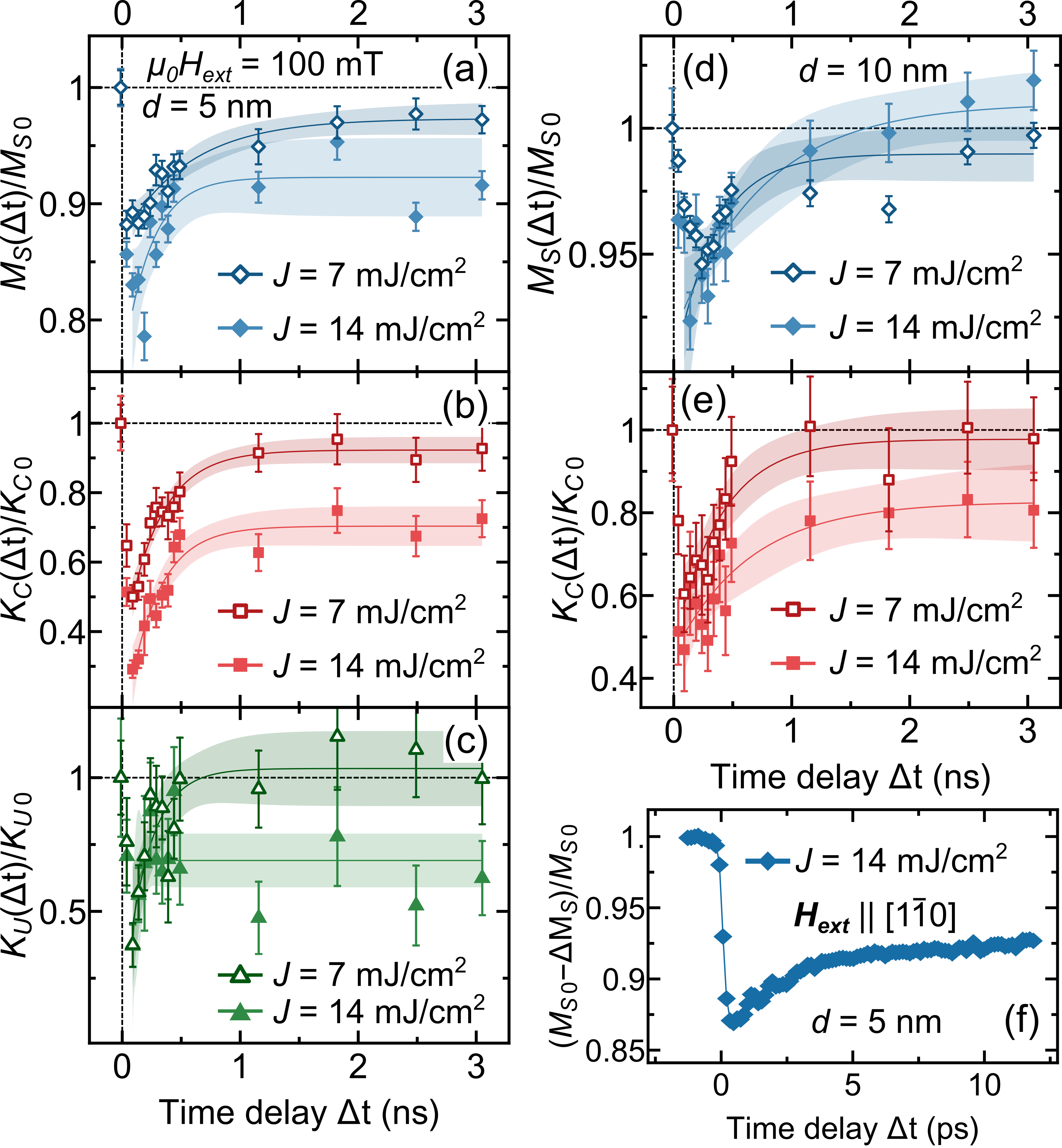}
\caption{\label{fig:Kc_Ku_vs_t} 
(a)-(e) Time dependencies of the magnetic parameters vs pump-probe delay time $\Delta t$ obtained from the LMOKE field dependencies~(Fig.~\ref{fig:loop}) using Eqs.~(\ref{eq:K_vs_t}). 
The lines show the fit according to Eq.~(\ref{eq:kvst}).
\textcolor{newtext}{Shaded areas show 95\,\% confidence levels for the fit.}
(f) Laser-induced demagnetization as measured using the conventional pump-probe technique in the 5-nm-thick film.
}
\end{figure}

\noindent Figure~\ref{fig:Kc_Ku_vs_t}\,(a)-\ref{fig:Kc_Ku_vs_t}\,(e) shows the temporal dependencies $M_S(\Delta t)$, $K_C(\Delta t)$, and $K_U(\Delta t)$, obtained from Eqs.~(\ref{eq:K_vs_t}) using the parameters of the hysteresis loops. 
Only the time dependencies $H_A(\Delta t)$ differ for two HA; thus we use averaged $M_S(\Delta t)$ for both directions. 
The experimentally measured $K_U(\Delta t)$ could be obtained for the 5 nm-thick film only~[Fig.~\ref{fig:Kc_Ku_vs_t}\,(c)].
In galfenol films on GaAs substrates, the uniaxial anisotropy is substrate-induced and is more pronounced in thin films~\cite{farle-1998}.
On the other hand, cubic anisotropy is defined by the bulk material.
Thus, the changes $K_C(\Delta t)$ can be reliably obtained from the experimental data for both the 5 and 10\,nm-thick films, while $K_U(\Delta t)$ is determined for the former film only.

For all parameters, a rapid decrease in the absolute value is observed, followed by a slow relaxation. 
The recovery process is approximated using the exponential function:

\begin{equation} \label{eq:kvst}
X (\Delta t) = X_0 - \Delta X e^{-\Delta t/\tau},
\end{equation}

\noindent where $X$ is $M_S$, $K_C$, or $K_U$, $\Delta X$ is the laser-induced change of the corresponding parameter, and $\tau$ is a relaxation time.
Fits of the experimental data with Eq.~(\ref{eq:kvst}) are shown by the lines in Fig~\ref{fig:Kc_Ku_vs_t}.

Abrupt reduction of $M_S$ is ascribed to ultrafast demagnetization \cite{Bigot:PRL1996}.
This is confirmed by the time dependence $M_S(\Delta t)$ measured at saturation in the same LMOKE geometry using the conventional pump-probe technique~[Fig.~\ref{fig:Kc_Ku_vs_t}\,(f)]. 
As a result of laser excitation, $M_S$ reduces from the initial value to a minimum with a characteristic time less than 1\,ps.
In agreement with literature data~\cite{roth2012Temperature,koopmans2010explaining} the initial partial recovery of $M_S$ occurs within a few ps, followed by a slower recovery seen in Fig.~\ref{fig:Kc_Ku_vs_t}\,(a) and \ref{fig:Kc_Ku_vs_t}\,(d).
It is established that after initial recovery, spins, electrons, and lattice are in equilibrium with each other~\cite{koopmans2010explaining}.
Below we focus our discussion on time delays exceeding this range.

Values of relaxation times $\tau$ for $K_C$, $K_U$, and $M_S$ for the two studied films and two laser pulse fluences are given in Table~\ref{table:relaxation_times_KcKuMs}. 
All parameters show partial or full recovery with characteristic times $\tau$ of several hundreds of picoseconds [see Fig.~\ref{fig:Kc_Ku_vs_t}\,(a)-\ref{fig:Kc_Ku_vs_t}\,(e)]. 
At time delays longer than 1\,ns the relaxation time becomes much longer, and $M_S$, $K_C$, and $K_U$ can be considered as constants at this timescale.
At $J=$ 14\,mJ/cm$^2$, the material parameters do not fully recover to their initial values within 3\,ns. 
At $J=$ 7\,mJ/cm$^2$, the values of the material parameters at $\Delta t=$ 3\,ns are close to those before the excitation.
The relaxation process longer than 3\,ns can be used to locally control the parameters of propagating spin waves in future opto-magnonic devices~\cite{VogelNPhys:2015, dzyapko2016reconfigurable, Filatov_spectrum_2020_JPCS}. 
On the other hand, complete relaxation of magnetic parameters in less than 1\,ns is beneficial for magnetization switching processes in information storage devices~\cite{Davies-PRL2019, davies2020pathways}.

\begin{table}
\caption{\label{table:relaxation_times_KcKuMs}
Relaxation time $\tau$ (ps) of the saturation magnetization $M_S$ and anisotropy parameters $K_C$ and $K_U$ extracted from the fit of the LMOKE field dependencies (Fig.~\ref{fig:loop}) using Eq.~(\ref{eq:kvst}) for the 5-nm and 10-nm films and the pump fluences 7 and 14\,mJ/cm$^2$.
}
\begin{ruledtabular}
\begin{tabular}{c|ccc|cc}
$J$ & $K_C$ & $K_U$ & $M_S$ & $K_C$ & $M_S$ \\
(mJ/cm$^2$) & \multicolumn{3}{c|}{$\tau$ (ps) for $d=$5\,nm} & \multicolumn{2}{c}{$\tau$ (ps) for $d=$10\,nm}\\
\colrule
 7 & 300 $\pm$ 50 & 210 $\pm$ 80 & 560 $\pm$ 150 & 400 $\pm$ 120 & 340 $\pm$ 140\\ 
 14 & 270 $\pm$ 40 & 40 $\pm$ 40 & 230 $\pm$ 110 & 640 $\pm$ 330 & 740 $\pm$ 290\\ 
\end{tabular}
\end{ruledtabular}
\end{table}

We note that the obtained relaxation dynamics of the anisotropy parameters at thermal equilibrium between electrons, spins and the lattice has a form similar to the one obtained from magnetization precession of a thin Fe film~\cite{carpene-PRB2010}.

\begin{figure}
\includegraphics[width=0.8 \linewidth]{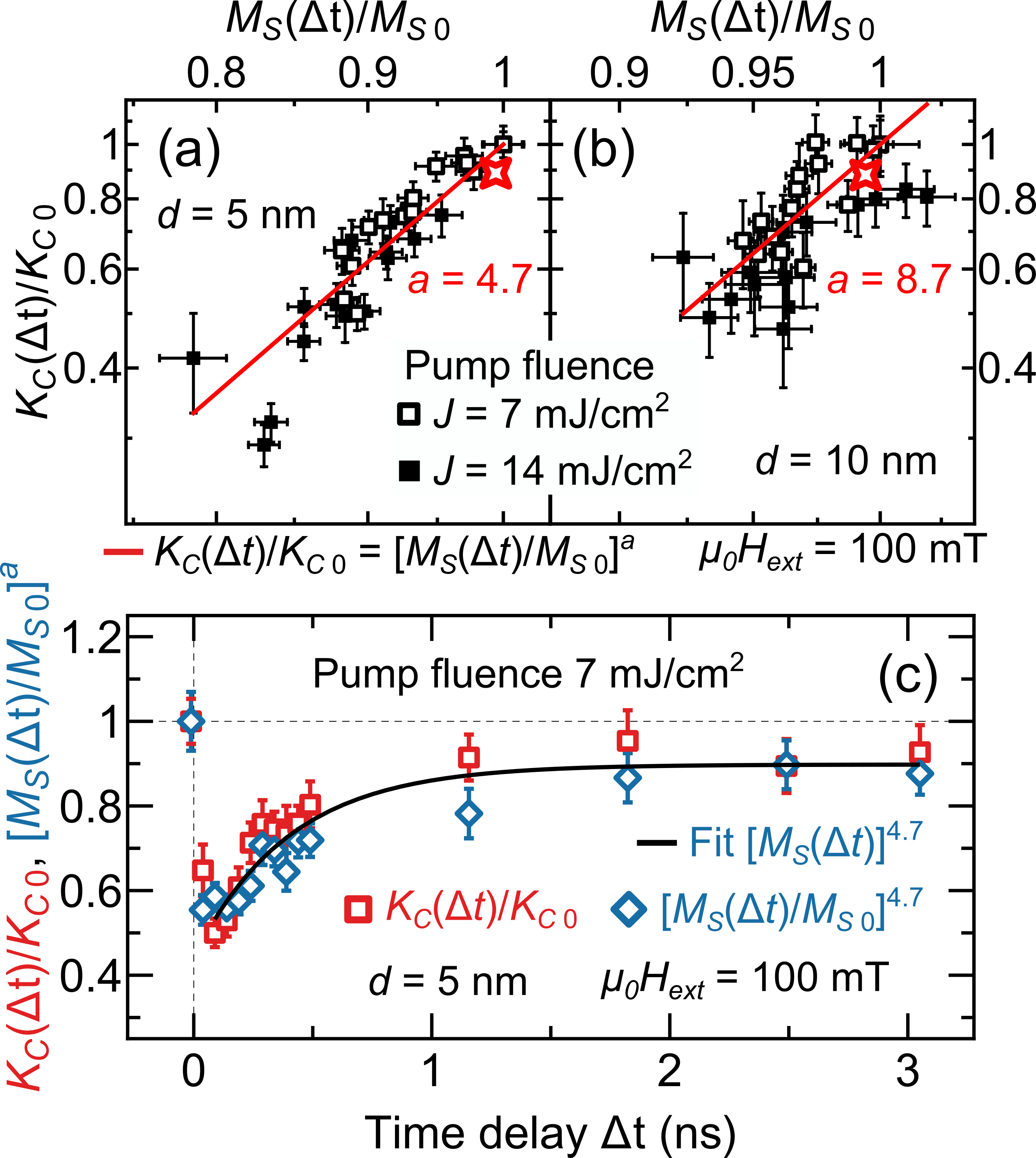}
\caption{\label{fig:PowerLaw}
(a), (b) Logarithmic plot of the cubic anisotropy parameter $K_C$ vs saturation magnetization $M_S$ . 
Power-law fit is shown by the solid line.
The best fit corresponds to $a = 4.7$ and 8.7 for film thickness $d=$ 5 and 10\,nm, respectively.
(c) Illustration of the applicability of the power law for the case of temporal evolution at a pump fluence 7\,mJ/cm$^2$.
Film thickness $d=$ 5\,nm, $\mu_0 H_{ext} = 100$\,mT.
Red stars at panels (a) and (b) are $K_C$ and $M_S$ obtained from the analysis of precession (Table\,\ref{table:Parameters_FFT_windows}) with $K_{C0}=K^c_C$ and $M_{S0}=M^c_{S}$.}
\end{figure}

Separation of the dynamics of $M_S(\Delta t)$ and $K_C(\Delta t)$ allow us to verify the applicability of a power-law dependence between these parameters, established for the thermal equilibrium in~\cite{Zener-PRB1954}:

\begin{equation} \label{eq:PowerLaw}
K_C(\Delta t)/K_{C 0} = [M_S(\Delta t)/M_{S 0}]^a,
\end{equation}

\noindent where $K_{C 0}$ and $M_{S 0}$ are the cubic anisotropy parameter and saturation magnetization values before the excitation, respectively. 
The exponent $a$ is determined only by the origin of the anisotropy and for the single-ion cubic one is expected to be equal to 10. 

In Fig.~\ref{fig:PowerLaw}\,(a) and \ref{fig:PowerLaw}\,(b) we plot $K_C / K_{C 0}$ vs $M_S / M_{S 0}$ as obtained at different time delays $\Delta t$. 
Fitting of these dependencies with Eq.~(\ref{eq:PowerLaw}) gives $a=4.7 \pm 0.2$ and $8.7 \pm 1$ for the 5 and 10\,nm-thick films, respectively.
For the 10\,nm-thick film, the change in $M_S$ is small, which increases the relative error in determining the change in the cubic anisotropy parameter and the exponent. 
However, as can be seen in Fig.~\ref{fig:PowerLaw}\,(a) for the 5\,nm-thick film, the power law [Eq.~(\ref{eq:PowerLaw})] describes well the changes in the material parameters at both pump pulse fluences and at all $\Delta t$, apart from the case of the largest $\Delta M_S$ and $\Delta K_C$, corresponding to the short-time delays.
To demonstrate this we plot in Fig.~\ref{fig:PowerLaw}\,(c) $K_C(\Delta t)/K_{C0}$ and $[M_S(\Delta t)/M_{S0}]^{4.7}$ for the 5\,nm thick film. 
The deviation of the relation between $\Delta K_C$ and $\Delta M_S$ from the power law with exponent~(10) is more pronounced for the thinner film.
This feature can be explained by an additional two-site contribution to the anisotropy present along with the conventional single-ion one, as recently noted in~\cite{evans2020temperature}.

Thus, our results show that at the timescale spanning from several picoseconds up to 3\,ns in a magnetic metal, the evolution of the saturation magnetization and cubic anisotropy parameter do indeed relate to each other according to the power law known for the case of thermal equilibrium. 
We note that experimentally the temporal evolution of the saturation magnetization $M_S$ can be readily obtained using the conventional pump-probe technique [see, e.g., Fig.~\ref{fig:Kc_Ku_vs_t}\,(f)]. 
The evolution of the magnetic anisotropy parameters cannot be obtained with matching precision, as it requires the analysis of hysteresis loops.
Therefore, the verified power-law relation between the time dependencies of $M_S$ and $K_C$ provides an approach for the analysis of the evolution of magnetic parameters in this important time range. 
We demonstrate this in the following section for laser-induced magnetization precession in the studied films, which is observed at the considered timescale.

\subsection{\label{Sec:EffectOfEvolution} Effect of evolution of anisotropy parameters on magnetization precession}

\begin{figure}
\includegraphics[width=0.8 \linewidth]{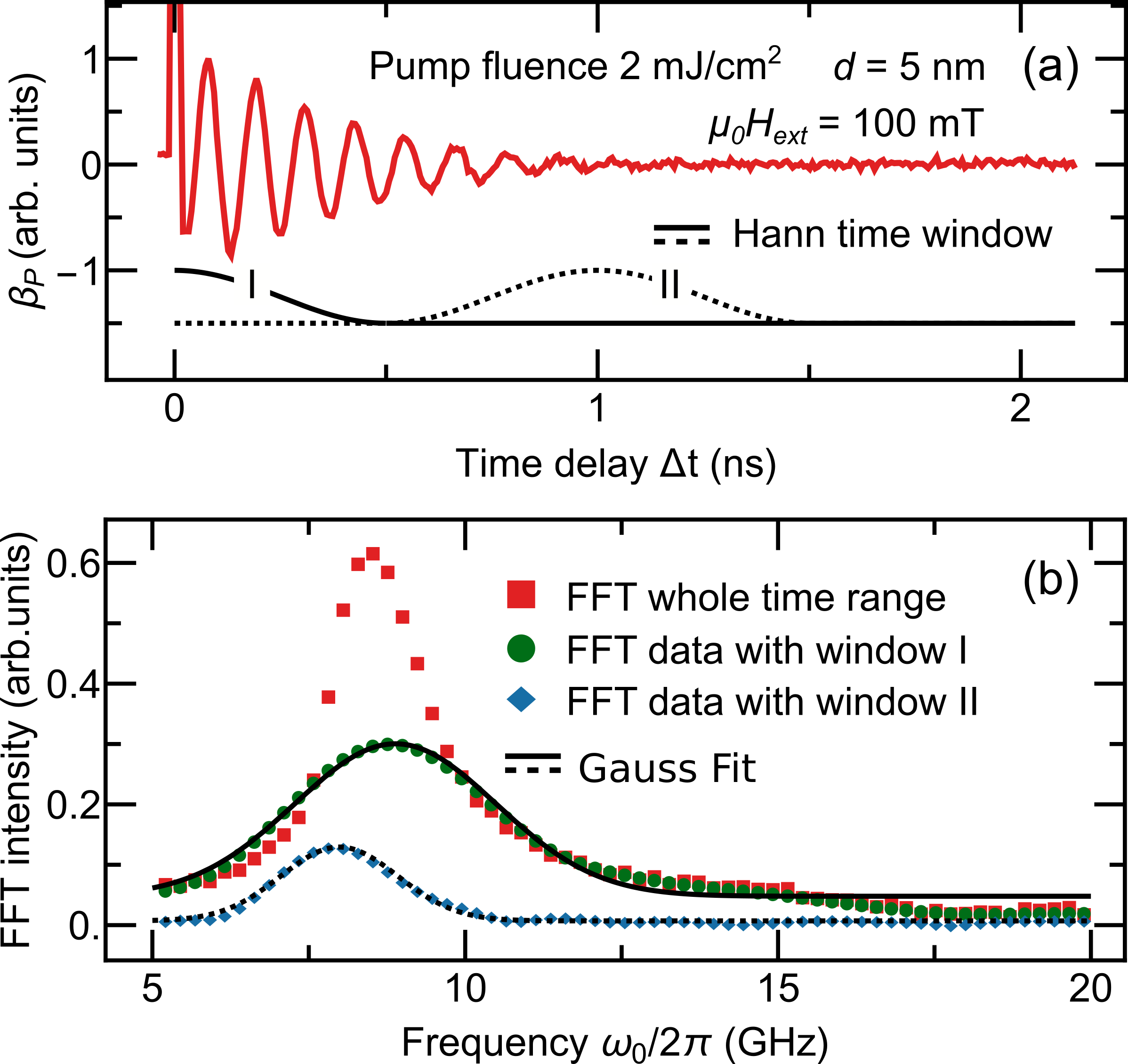}
\caption{\label{fig:FFT_vs_t} 
(a) Experimental TR-PMOKE signal (red line). 
Lines I and II show the applied Hann time windows.
(b) Whole-time FFT (red squares) and short-time FFT (green and blue diamonds) of data from (a). 
Lines show Gaussian fits of data. }
\end{figure}

To reveal the effect of the temporal evolution of $K_U$, $K_C$, and $M_S$ on the laser-induced magnetization precession, we performed experiments with the optical excitation and detection of the precession in the TR-PMOKE geometry [Fig.~\ref{fig:geometry}\,(b)].
\textcolor{newtext}{We use pump pulses of 2~mJ/cm$^2$ fluence which ensure that $M_S$, $K_C$, and $K_U$ restore values close to the initial ones after $\sim$1~ns [see Fig.~\ref{fig:Kc_Ku_vs_t}]}
An example of a measured signal for the 5~nm thick film with $\mathbf{H_{ext}}$ directed along $[1\overline{1}0]$ is shown in Fig.~\ref{fig:FFT_vs_t}\,(a).
In the frequency spectrum of the precession, obtained by the fast Fourier transform~(FFT) in the whole experimental time range 0-2~ns, a single strongly asymmetric peak is observed [squares in Fig.~\ref{fig:FFT_vs_t}\,(b)].
To elucidate the frequency composition of the FFT peak, we can now use the information on the relaxation timescales for the magnetization and anisotropy parameters discussed above. 
A strict approach would require modeling the magnetization precession around the time-dependent effective field.
Instead, in order to use a simple approach based on FFT, we divided the experimental timescale into two time ranges corresponding to the range of the transient changes of the magnetic parameters following the excitation~(I), and the range of parameters corresponding to the cooled film~(II) [see Fig.~\ref{fig:Kc_Ku_vs_t} and Table~\ref{table:relaxation_times_KcKuMs}].
Then we used a short-time FFT with a Hann window~\cite{zierhofer2007data} with the width of 1\,ns at the central positions of $\Delta t = 0$ (range~I) and 1\,ns (range~II), as shown by black lines in Fig.~\ref{fig:FFT_vs_t}\,(a).
The operation yields two peaks in the spectrum approximated by Gaussian functions with different central frequencies: \textcolor{newtext}{8.88 $\pm$ 0.03\,GHz and 7.99 $\pm$ 0.02\,GHz in the time windows I and II, respectively} [lines in Fig.~\ref{fig:FFT_vs_t}\,(b)].

\begin{figure*}
\includegraphics[width=0.65 \linewidth]{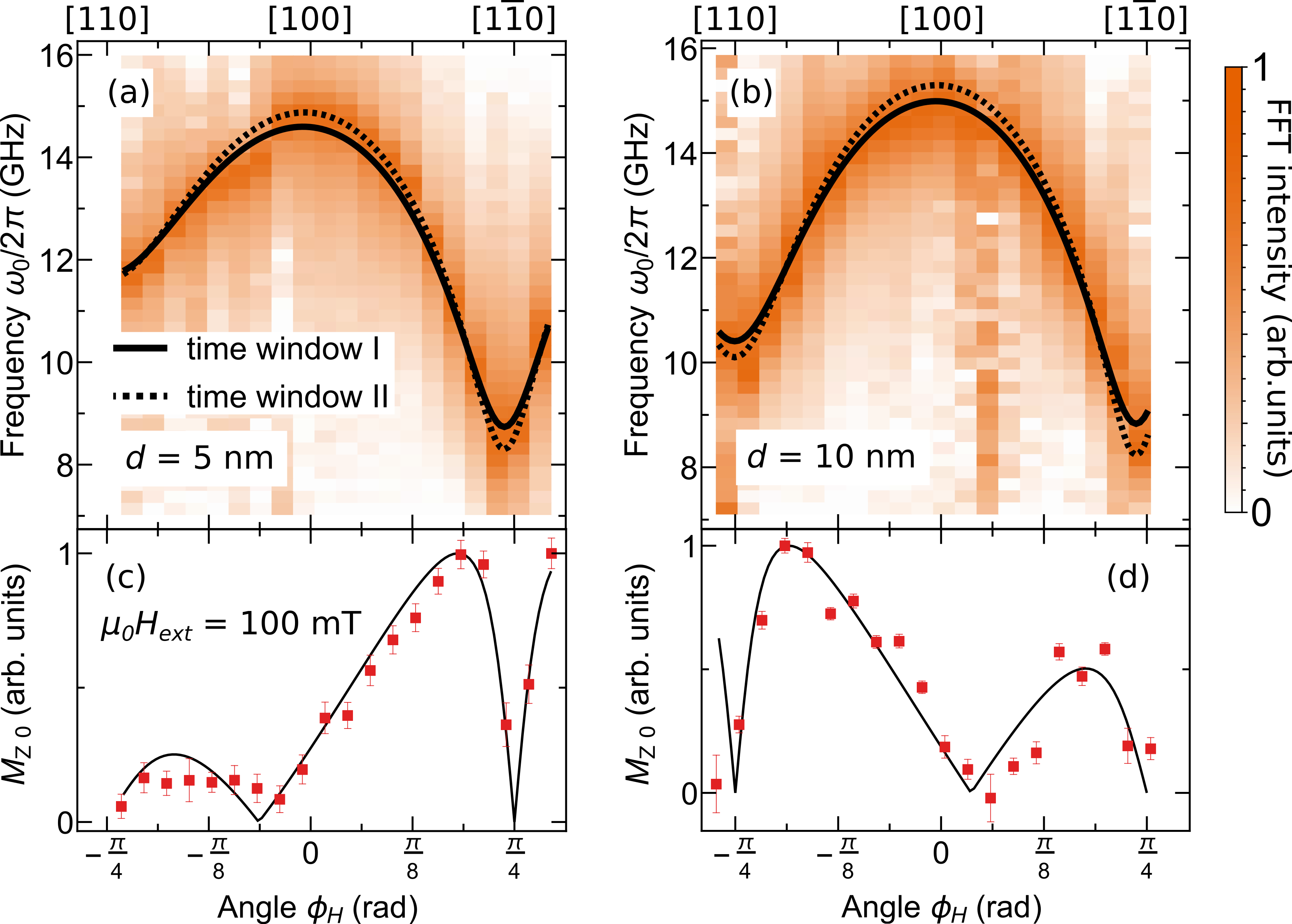}
\caption{\label{fig:azimuthalDep} 
(a), (b) Azimuthal dependencies of whole-time FFTs of the detected magnetization precession signals.
Lines show the results of approximations of the azimuthal dependencies of the frequencies using Eq.~\ref{eq:omega0} for the short-time FFTs in the windows I and II (see Fig~\ref{fig:FFT_vs_t}).
(c), (d) Azimuthal dependencies of the precession amplitudes.
}
\end{figure*}

Figures~\ref{fig:azimuthalDep}\,(a) and \ref{fig:azimuthalDep}\,(b) show the azimuthal dependencies of the frequencies of the two peaks in the FFT spectra obtained at various orientations $\phi_H$ of $\bf H_{ext}$ in the range between $[1\overline{1}0]$ and $[110]$.
The azimuthal dependence of the precession frequency reflects the cubic and uniaxial anisotropy of the film.
The difference between the precession frequency in the $[110]$ and $[1\overline{1}0]$ directions is proportional to $K_U$ and increases with decreasing film thickness $d$. 

The magnetic anisotropy parameters are determined from the approximation of the azimuthal dependencies of the central positions of peaks I and II~[Fig.~\ref{fig:FFT_vs_t}\,(b)] (see Appendix~\ref{Sec:AppendixPrecession} for details).
The approximation results are shown in Fig.~\ref{fig:azimuthalDep}\,(a) and \ref{fig:azimuthalDep}\,(b) by lines.
The corresponding magnetic parameters of the film right after excitation, i.e. within the FFT window I ($M_S^h, K_C^h, K_U^h$), and after cooling down, i.e., in the FFT window II ($M_S^c, K_C^c, K_U^c$), are shown in the Table~\ref{table:Parameters_FFT_windows}. 
The values $K_C^h/K_C^c$ and $M_S^h/M_S^c$ are also added to the plots $K_C/K_{C0}$ vs $M_S/M_{S0}$ [Fig.~\ref{fig:PowerLaw}\,(a) and \ref{fig:PowerLaw}\,(b)], and are close to the power-law relations deduced in Sec.~\ref{Sec:TemporalEvolution} for the corresponding films.

Figures~\ref{fig:azimuthalDep}\,(c) and \ref{fig:azimuthalDep}\,(d) show the azimuthal dependencies of the out-of-plane component of the magnetization $M_{z0}$, characterizing the precession amplitude and found as the area under the FFT curve obtained with the time window I.
As can be seen, in both studied films the precession amplitude depends on the applied field direction, with larger amplitudes seen when the field is applied along neither EA nor HA. 
However the characteristics of the dependence noticeably change with the change of the film thickness.

\begin{table}
\caption{\label{table:Parameters_FFT_windows}%
Magnetic parameters of the film right after excitation ($M_S^h, K_C^h, K_U^h$) and after cooling down ($M_S^c, K_C^c, K_U^c$)
}
\begin{ruledtabular}
\begin{tabular}{c|cc|cc|cc}
 d & $\mu_0M_S^h$  & $\mu_0M_S^c$  & $K_C^h$ & $K_C^c$ & $K_U^h$ & $K_U^c$ \\
 (nm) & \multicolumn{2}{c|}{(T)} & \multicolumn{2}{c|}{($10^4 J/m^3$)} & \multicolumn{2}{c}{($10^4 J/m^3$)} \\
\colrule
 5 & 1.69 & 1.7 & 2.4 & 2.7 & -1.4 & -1.5 \\ 
 10 & 1.69 & 1.7 & 3 & 3.4 & -0.7 & -0.8 \\ 
\end{tabular}
\end{ruledtabular}
\end{table}

Having determined the magnetic parameters of the films right after excitation and after their recovery, we now turn to describing the azimuthal dependencies of the laser-induced precession amplitude [Fig.~\ref{fig:azimuthalDep}\,(c) and \ref{fig:azimuthalDep}\,(d)].
The laser-induced heating alters the equilibrium direction of ${\bf M}$ described by angle $\phi_0$ on a timescale much shorter than the precession period. 
Therefore, as detailed in~\cite{khokhlov-PRAppl2019}, the precession amplitude is proportional to the pump-induced change of $\phi_0$.
As noted above, the magnetic parameters completely relax to their initial values in a time of the order of 1\,ns at $J <$ 7\,mJ/cm$^2$ (see Fig.~\ref{fig:Kc_Ku_vs_t}). 
Therefore, one can use the values of the magnetic parameters obtained from the approximation of the azimuthal dependencies of $\omega_0$ for $\Delta t =0$ and 1\,ns, i.e., in the FFT windows I and II, to determine $\phi_0$ for the heated (${\phi_0}^h$) and cooled (${\phi_0}^c$) film, respectively. 
\textcolor{newtext}{These values serve as a good approximation of the effective field's orientation before (${\phi_0}^c$)  and immediately after the excitation (${\phi_0}^h$).}
Thus, the \textcolor{newtext}{difference ${\phi_0}^h-{\phi_0}^h$} corresponds to the amplitude of the in-plane deviation of \textbf{M}, while the experimental data are obtained for the $M_z$ component. Therefore, it is necessary to take into account the ellipticity $N^h$ of the precession (see Appendix~\ref{Sec:AppendixPrecession} for details), found for the time window I.
Thus we get for $M_{z 0}$

\begin{equation} \label{eq:asw}
M_{z 0} = M_S \sin{[N^{h} ({\phi_0}^{h} - {\phi_0}^{c})]}.
\end{equation}

Approximation of the azimuthal dependencies of the precession amplitude using~Eq.~(\ref{eq:asw}) is shown in Fig.~\ref{fig:azimuthalDep}\,(c) and \ref{fig:azimuthalDep}\,(d) by lines. 
The approximation is carried out up to a scaling factor and demonstrates good agreement with the experimental data. 
The difference between the azimuthal dependencies of the precession amplitudes for the 5 and 10\,nm-thick films evident in Fig.~\ref{fig:azimuthalDep}\,(c) and \ref{fig:azimuthalDep}\,(d) stems from different relationships between the cubic and uniaxial anisotropy parameters.
Asymmetry of azimuthal dependencies of $\omega_0$ and $M_{z 0}$ in the investigated range of $\phi_H$ depends on the growth-induced uniaxial anisotropy parameter and is more pronounced for the 5\,nm-thick film. 

\section{Conclusions}

We have studied the laser-induced dynamics of magnetization and magnetic anisotropy in thin films of the ferromagnetic metal galfenol based on the analysis of the evolution of the hysteresis loops following ultrafast laser-induced heating.
We have demonstrated that the abrupt decrease of magnetization and magnetic anisotropy parameters is followed by an exponential recovery occurring within a nanosecond and being slower for the thicker film. 
We verified that the power-law relation between the magnetization and magnetic anisotropy known for the thermal equilibrium holds during the laser-induced relaxation processes over the time range from a few picoseconds up to 3~nanoseconds. 
This suggests that the temporal evolution of the magnetic anisotropy in the metallic films excited by femtosecond laser pulses can be obtained from the laser-induced demagnetization dynamics, which, in turn, can be directly measured in pump-probe experiments.
We note that magnetization recovery following ultrafast demagnetization proceeding at pico-, nano-, and even microsecond timescales recently became a subject of extensive research motivated by the practical importance of such timescales~\cite{Wang:PRRes2021}.

Indeed, understanding the evolution of magnetic anisotropy along with magnetization at pico- and nanosecond timescales allows more detailed evaluation of the dynamics of magnetization excited by laser pulses.
We demonstrate it for the case of laser-driven magnetization precession in the studied films which is observed at the timescale comparable to the characteristic time of the magnetization and magnetic anisotropy relaxation.
By taking into account the timescales at which the recovery of the magnetic parameters occurs, we unveil a shift of the precession frequency of $\sim$\,0.9\,GHz (10\,\% of frequency) occurring within a nanosecond after the excitation.

Such time-dependent shifts of precession frequency controlled by laser-induced anisotropy changes may be further exploited in optically reconfigurable magnonic devices based on tuning the magnetostatic wave dispersion~\cite{VogelNPhys:2015}.
The possibility to tune the eigenfrequency of magnetic dynamics may find its application in the realization of laser-assisted spin-torque nano-oscillators~\cite{Farkhani:Neuro2020}.

Finally, we note that, for optically reconfigurable magnonics~\cite{VogelNPhys:2015}, plane and patterned thin galfenol films on GaAs substrates are found to be prospective structures. 
Ultrafast laser-induced heating of galfenol resulting in demagnetization and anisotropy changes has been shown to trigger magnetization precession~\cite{kats2016ultrafast}, propagating magnetostatic~\cite{khokhlov-PRAppl2019, Filatov_spectrum_2020_JPCS} and standing spin~\cite{Scherbakov:PhysRevAppl2019} waves, coupled magnon-phonon modes~\cite{godejohann2020Magnon}, and spin currents~\cite{DanilovPRB:2018}. 
The relaxation processes investigated here may both facilitate and hinder such processes. 

\begin{acknowledgments}
The work of P.I.G., D.V.K., Ia.A.F. and N.E.Kh. was funded by RFBR (Grant No.~20-32-70149).
L.A.Sh. acknowledges the Foundation for the Advancement of Theoretical Physics and Mathematics “BASIS”(Grant No.~19-1-5-101-1).
A.M.K. acknowledges RSF (Grant No.~20-12-00309).
\end{acknowledgments}

\appendix

\section{\label{Sec:AppendixLoop} \textcolor{newtext}{Fitting the hysteresis loops}}

\textcolor{newtext}{
In Eq.~(\ref{eq:Hist_fit}), we use a second-order polynomial $F(x)$ with coefficients $\{ a_1, a_2 \}$ to fit the experimental hysteresis loops.
The first-order polynomial does not provide a good fit in the case of the 10\,nm thick film.
Using a third-order polynomial leads to the zeroing of the coefficient before the cubic term. 
Nonlinear growth of the signal at $|H_{ext}|<H_A$ found for the 10~nm-thick film suggests that there was a small misalignment between the applied field and the hard axes. 
The polynomial coefficients $\{ a_1, a_2 \}$ used in Eq.~\ref{eq:Hist_fit} for two magnetization directions of $[1\overline{1}0]$ and $[110]$ are $\{ 6.54\,10^{-3}, 8.60\,10^{-6} \}_{5\,nm}$ along $[1\overline{1}0]$, $\{3.87\,10^{-3}, 8.72\,10^{-6} \}_{5\,nm}$ along [110] and $\{-1.35\,10^{-9}, 5.37\,10^{-5} \}_{10\,nm}$ along $[1\overline{1}0]$, $\{-1.50\,10^{-9}, 5.95\,10^{-5} \}_{10\,nm}$ along [110] for 5 and 10\,nm-thick films, respectively. 
The average error in determining the $H_A$ field using Eq.~(\ref{eq:Hist_fit}) is 3\,mT and 5\,mT for 5 and 10\,nm thick films, respectively. 
}

\section{\label{Sec:AppendixPrecession} \textcolor{newtext}{Analysis of azimuthal dependencies of the precession parameters}}

\begin{figure}
\includegraphics[width=0.92 \linewidth]{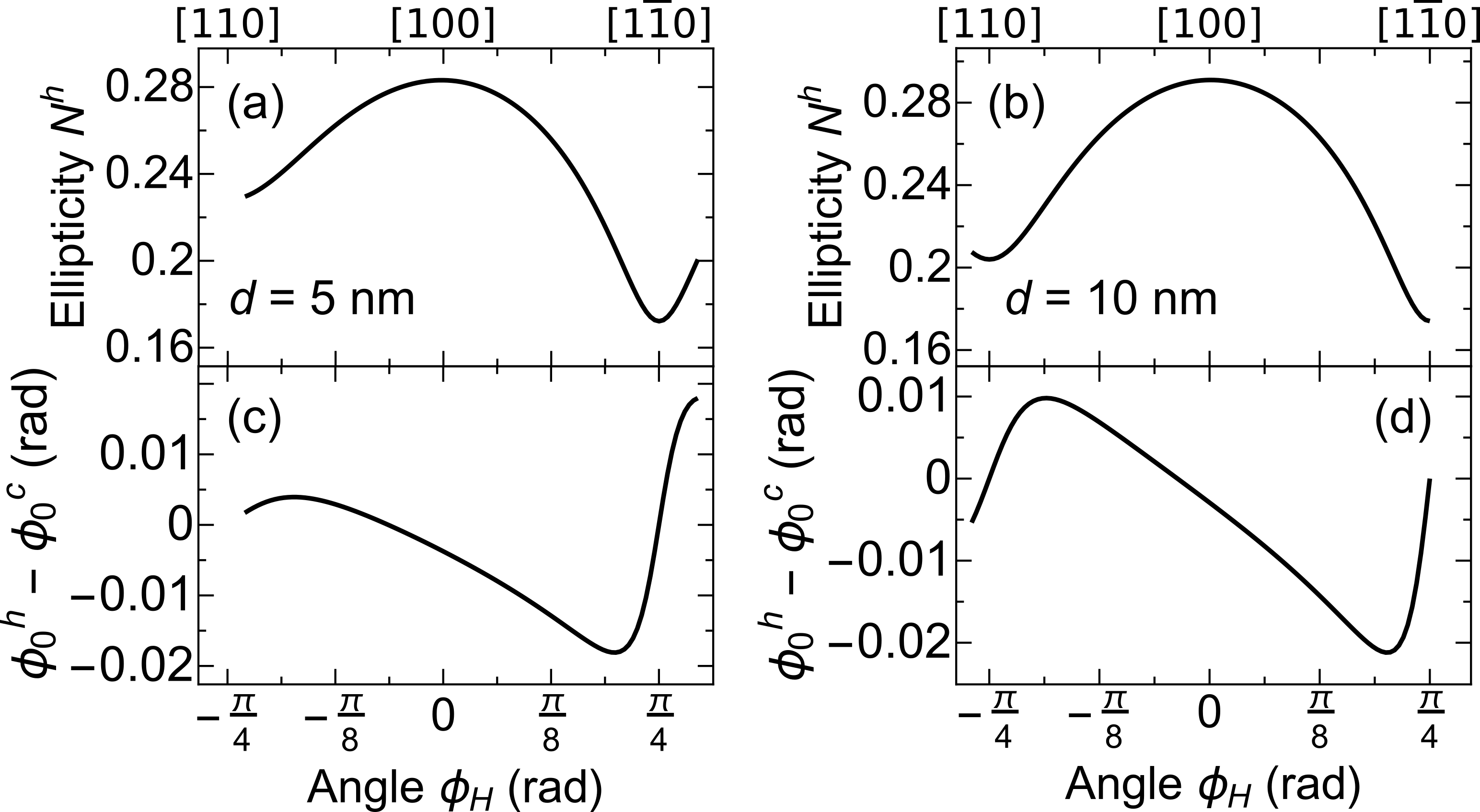}
\caption{\label{fig:dPhi_and_N_vs_PhiH} 
(a), (b) Ellipticity of magnetization precession after ultrafast heating for 5 and 10\,nm-thick galfenol films.
(c), (d) Difference between equilibrium magnetization directions for heated and cooled 5 and 10-nm-thick galfenol films.  
}
\end{figure}

We use the Smit-Beljers approach to describe the azimuthal dependencies of the precession parameters~\cite{smit1955ferromagnetic}.
Taking into account a small deviation of the magnetization from its equilibrium orientation
and performing calculations similar to~\cite{gurevich1996magnetization}, we obtain expressions for the frequency $\omega$ and ellipticity $N$ of the laser-induced magnetization precession:
\vspace{1mm} 
\begin{eqnarray}
\label{eq:omega0}
 & \omega_0 &= \cfrac{\gamma}{M_S \sin{\theta_0}} \sqrt{U_{\theta \theta} U_{\phi \phi} - {U_{\theta \phi}}^2},\\
\label{eq:ellipticity}
 & N &= \cfrac{\sqrt{U_{\theta \theta} U_{\phi \phi} - {U_{\theta \phi}}^2}}{U_{\theta \theta} - \alpha U_{\theta \phi} \csc{\theta_0}}
\end{eqnarray}
\vspace{1mm} 
\noindent where $\gamma$ is the gyromagnetic ratio, $U_{i j} = \cfrac{\partial^2 U}{\partial i \partial j}, \{i, j\} = \{\theta, \theta\},\ \{\phi, \phi\}\ \rm{and}\ \{\theta, \phi\}$ at equilibrium direction of $\textbf{M}$ ($\theta = \theta_0$ and $\phi = \phi_0$). 
$U$ is the film free energy density:
\vspace{1mm} 
\begin{multline} \label{eq:U}
U = 
- \mu_0 H_{ext} M_S \sin{\theta} \cos{(\phi - \phi_H)} \\
+ \frac{1}{2} \mu_0 {M_S}^2 \cos^2\theta - \frac{K_U}{2} \sin{2 \phi} \sin^2\theta \\
+ \frac{K_C}{4} \big[\sin^4{\theta} \sin^2{2 \phi} + \sin^2{2 \theta}\big].
\end{multline}
\vspace{1mm} 
\noindent The polar angle $\theta_H$ of ${\bf H_{ext}}$ is equal to $\pi/2$ in our case.
\vspace{1mm} 
The azimuthal dependencies of the ellipticity $N^h$ right after excitation are shown in Fig.~\ref{fig:dPhi_and_N_vs_PhiH}\,(a,~b) for film thickness $d = 5$ and 10\,nm, respectively.
\vspace{1mm} 
\vspace{1mm} 
Due to the strong shape anisotropy and $\theta_H = \pi/2$, $\theta_0$ is always equal to $\pi/2$.
The $\phi_0$ is obtained from the solution of the equation for the equilibrium direction of $\textbf{M}$:
\vspace{1mm} 
\begin{equation}
\label{eq:dUdphi}
\cfrac{\partial U}{\partial \phi}\Big|_{\theta = \frac{\pi}{2}, \phi = \phi_0} = 0.
\end{equation}

The analytical solution for $\phi_0$ can be obtain in the case of $H_{ext} \gg H_{A}$.
However, a high value of $H_{ext}$ leads to a vanishing magnetization precession amplitude and cannot be used in the experiments.
Due to the that, we perform all experiments at $H_{ext} \sim H_{A}$ and determine $\phi_0$ numerically from Eq.~(\ref{eq:dUdphi}) for each direction of $H_{ext}$.

The azimuthal dependencies of the difference between the equilibrium magnetization direction $\phi_0$ for the heated (${\phi_0}^h$) and cooled (${\phi_0}^c$) film, i.e., in-plane angular amplitude, are shown in Fig.~\ref{fig:dPhi_and_N_vs_PhiH}\,(c) and \ref{fig:dPhi_and_N_vs_PhiH}\,(d) for film thickness $d = 5$ and 10\,nm, respectively.
The parameters of the material right after excitation and after complete cooling are determined from the azimuthal dependencies of the magnetization precession (Sec.~\ref{Sec:EffectOfEvolution}) using Eq.~\ref{eq:omega0}.

\normalem
\bibliography{References}

\end{document}